\documentclass{midl} 

\usepackage{mwe} 
\usepackage{multirow}
\jmlrvolume{}
\jmlryear{}
\jmlrworkshop{Extended Abstract -- MIDL 2019}

\title[CNN-Based Whole-Heart Segmentation in Non-Contrast CT]{CNN-Based Segmentation of the Cardiac Chambers and Great Vessels in Non-Contrast-Enhanced Cardiac CT}

\midlauthor{\Name{Steffen Bruns\nametag{$^{1}$}} \Email{s.bruns@umcutrecht.nl}\\
\Name{Jelmer M. Wolterink\nametag{$^{1}$}} \Email{j.m.wolterink@umcutrecht.nl}\\
\Name{Robbert W. van Hamersvelt\nametag{$^{2}$}} \Email{r.w.vanhamersvelt-3@umcutrecht.nl}\\
\Name{Tim Leiner\nametag{$^{2}$}} \Email{t.leiner@umcutrecht.nl}\\
\Name{Ivana I\v{s}gum\nametag{$^{1}$}} \Email{i.isgum@umcutrecht.nl}\\
\addr $^{1}$ Image Sciences Institute, UMC Utrecht, Heidelberglaan 100, 3584CX Utrecht, the Netherlands\\
\addr $^{2}$ Department of Radiology, UMC Utrecht, Heidelberglaan 100, 3584CX Utrecht, the Netherlands
}

\begin{document}

\maketitle

\begin{abstract}
    
\end{abstract}
Quantification of cardiac structures in non-contrast CT (NCCT) could improve cardiovascular risk stratification. However, setting a manual reference to train a fully convolutional network (FCN) for automatic segmentation of NCCT images is hardly feasible, and an FCN trained on coronary CT angiography (CCTA) images would not generalize to NCCT. Therefore, we propose to train an FCN with virtual non-contrast (VNC) images from a dual-layer detector CT scanner and a reference standard obtained on perfectly aligned CCTA images.

\section{Introduction}
Patients at risk of cardiovascular disease (CVD) may be identified based on accurate volumetric quantification of the cardiac structures~\cite{Nagarajarao,Narayanan}. Such quantification is feasible in coronary CT angiography~(CCTA)~\cite{Raman} and several automatic methods have been proposed to segment cardiac structures in CCTA~\cite{Ecabert,Zheng,Kirisli,Zhuang}.

However, many patients only undergo non-contrast CT~(NCCT) scanning~\cite{Sandfort}. It would thus be valuable to determine volumes of cardiac structures from the NCCT scan of these patients. Unfortunately, methods developed for segmentation in CCTA do not generalize to the NCCT domain. Recently,~\citet{Shahzad} have shown that segmentations of cardiac structures in the CCTA domain can be translated to the NCCT domain through multi-atlas segmentation. However, this approach requires a challenging inter-modality registration between CCTA and NCCT images. 

To address this, we propose to train a fully convolutional network~(FCN) for the segmentation of cardiac structures in NCCT using virtual-non-contrast (VNC) images~(\figureref{fig:Method}). VNC images mimic real NCCT images, but are reconstructed from a CT acquisition \textit{with} contrast-enhancement using a dual-layer detector CT scanner. Hence, a VNC image is perfectly aligned with the CCTA image of the same acquisition, and reference segmentations obtained in CCTA images can directly be used in the corresponding VNC images.

\section{Materials and methods}
\begin{figure}[tbp]
\floatconts
  {fig:Method}
  {\caption{Method overview. Left: A fully convolutional network (FCN) is trained with VNC images and CCTA segmentations. CCTA and VNC images are made in a single acquisition and thus, their segmentations are perfectly aligned. Right: The FCN automatically segments cardiac structures in both VNC and NCCT images.}
  }
  {\includegraphics[width=0.85\linewidth]{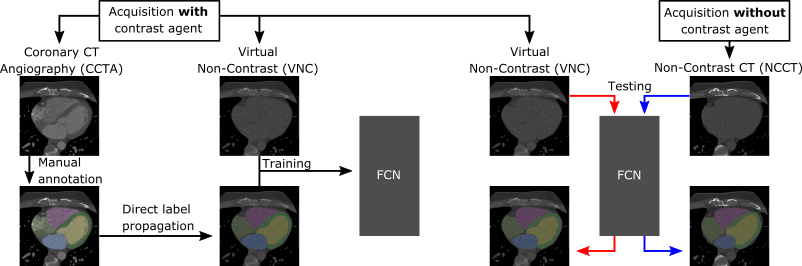}}
\end{figure}
We used a primary data set~\cite{Hamersvelt} consisting of CT images of 18 patients scanned on a Philips IQon dual-layer-detector CT scanner (Philips Healthcare, Best, The Netherlands). Hence, a CT acquisition with contrast enhancement could be reconstructed into a CCTA and a VNC image and these two images were perfectly aligned. Thus, reference segmentations of seven cardiac structures (left ventricular [LV] cavity, right ventricle, left atrium, right atrium, LV myocardium, ascending aorta, pulmonary artery trunk) obtained on CCTA images delineated the same structures on the VNC images. We also included a secondary data set consisting of 218 NCCT images with corresponding CCTA images to evaluate the capability of our method to segment NCCT images. These images were acquired on a Philips Brilliance iCT scanner (Philips Healthcare, Best, The Netherlands) without the option to reconstruct VNC images. All images had 0.34--0.56~$\text{mm}^2$ in-plane resolution, and 0.9~mm and 0.45~mm slice thickness and increment, respectively.

The FCN architecture we used is based on the 2D residual FCN proposed by ~\citet{Johnson}. We adapted the network to operate on 256$\times$256$\times$5 voxel 3D inputs and output 256$\times$256$\times$1 voxel predictions. We set the number of downsampling and upsampling layers to three, and the number of residual blocks to six. The FCN was trained with a mini-batch size of 32. Prior to training or testing, images were smoothed with a moderate Gaussian filter and resampled to an isotropic resolution of 0.8 $\times$ 0.8 $\times$ 0.8~$\text{mm}^3$. A connected component labeling was performed on obtained segmentations and for each structure, the largest connected component was retained in the final segmentation result.

The automatic segmentations on VNC images from the primary data set were evaluated quantitatively using the Dice similarity coefficient~(DSC) and the average symmetric surface distance~(ASSD). In the secondary data set, an expert inspected each NCCT image, the automatic NCCT segmentation, and the corresponding CCTA image, and assigned a grade to the segmentation based on the criteria proposed by~\citet{Abadi} where \textit{Grade 1} corresponds to very accurate segmentations and \textit{Grade 5} to a failed segmentation.


\section{Experiments and results}
We performed a six-fold nested cross-validation on the 18 VNC images from the primary data set. An ensemble of six models, one for each cross-validation fold, was used for testing on the secondary data set. Each model was trained for 10000 iterations using Adam optimization, an initial learning rate of 0.001, and a 70\% learning rate decay every 2000 iterations. The negative sum of soft Dice scores over all classes was used as the loss function.

\tableref{tab:DiceVNCPivot} lists quantitative results for the automatic segmentations in the 18 VNC images from the primary data set. Automatic segmentations of the 218 NCCT images in the secondary data set were qualitatively evaluated to assess the generalization of trained models to NCCT images. The expert assigned 27 segmentations~(12\%) to Grade 1~(very accurate), 119 segmentations~(55\%) to Grade 2, 42 segmentations~(19\%) to Grade 3, 15 segmentations~(7\%) to Grade 4, and 15 segmentations~(7\%) to Grade 5~(segmentation failed). Hence, most segmentations contained slight errors that are unlikely to significantly impact volume measurements. \figureref{fig:Qualitative} shows an automatic segmentation on an NCCT image.

\begin{table}[tbp]
\floatconts
  {tab:DiceVNCPivot}
  {\caption{Average ($\pm$ SD) per-class Dice similarity coefficients~(DSC) and average symmetric surface distances~(ASSD, in mm) between reference segmentations and automatic segmentations on virtual non-contrast (VNC) images for left ventricular cavity (LV-C) and myocardium (LV-M), right ventricle (RV), left atrium (LA), right atrium (RA), ascending aorta (AA), and pulmonary artery (PA) trunk.}}
{\resizebox{\linewidth}{!}{
\begin{tabular}{llllllll}
     & \textbf{LV-C}   & \textbf{LV-M}   & \textbf{RV}     & \textbf{LA}     & \textbf{RA}     & \textbf{AA}     & \textbf{PA}     \\
DSC  & $0.89 \pm 0.03$ & $0.84 \pm 0.04$ & $0.91 \pm 0.03$ & $0.92 \pm 0.01$ & $0.90 \pm 0.04$ & $0.94 \pm 0.01$ & $0.86 \pm 0.06$ \\
ASSD & $1.67 \pm 0.49$ & $1.15 \pm 0.27$ & $1.38 \pm 0.56$ & $1.19 \pm 0.35$ & $1.48 \pm 0.61$ & $0.75 \pm 0.24$ & $1.97 \pm 1.55$
\end{tabular}}}
\end{table}

\begin{figure}[tbp]
\floatconts
  {fig:Qualitative}
  {\caption{Automatic segmentation in a non-contrast CT image (secondary data set).}}
  {\includegraphics[width=\linewidth]{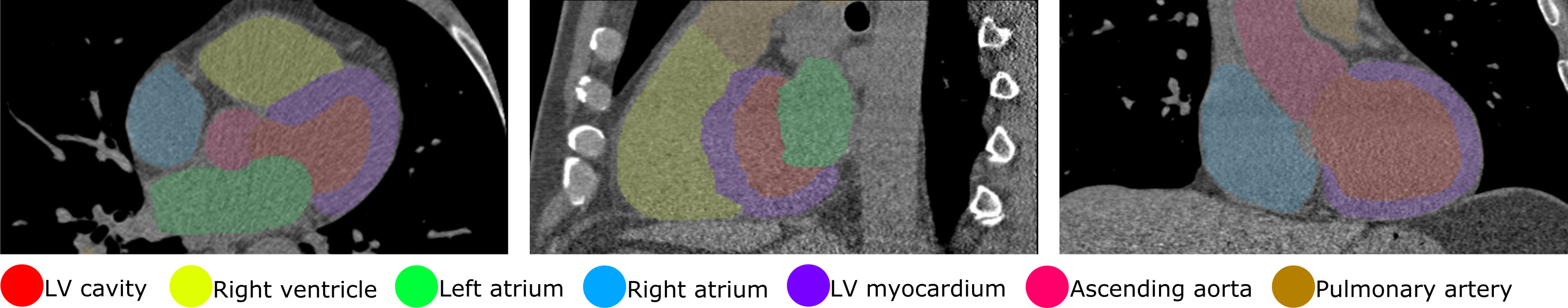}}
\end{figure}

\section{Discussion and conclusion}
We have presented an FCN that is able to segment seven cardiac structures in both VNC images and NCCT images. This allows accurate volume quantification of the cardiac chambers and great vessels even in the absence of contrast medium injection. Here, we addressed domain adaptation by using the physics properties of the dual-layer detector CT scanner, but future work may employ adversarial domain adaptation to translate images from the VNC domain to the NCCT domain~\cite{Lafarge}.


\midlacknowledgments{
This research is supported by the Dutch Technology Foundation STW, which is part of the Netherlands Organisation for Scientific Research (NWO) and partly funded by the Ministry of Economic Affairs and Philips Healthcare. We gratefully acknowledge the support of NVIDIA Corporation with the donation of the Titan Xp GPU used for this research.
}

\bibliography{midl-samplebibliography}

\end{document}